\documentstyle{article}

\def\QQa{\renewcommand{\baselinestretch}{1.3}\Huge\normalsize\large\small}

\begin{document}
\QQa

\large
\begin{flushright}
ITP.SB-92-71\\
Dec 8 ,1992
\end{flushright}

\begin{center}
\huge
Exotic Quantum Double,Its Universal R-matrix And Their Representations\\
\vspace{1cm}
\Large

Chang-Pu Sun \footnote {\large
Permanet address:Physics Dpartment,Northeast Normal University,
Changchun 130024,P.R.China}\\

Institute for Theoretical Physics,State University of New York,Stony Brook,
NY 11794-3840,USA\\

\vspace{1cm}
\huge
Abstract\\
\end{center}

\large
The exotic quantum double and its universal R-matrix for quantum Yang-Baxter
equation are constructed in terms
of Drinfeld's quantum double theory.As a new quasi-triangular Hopf algebra,
it is much different from those standard quantum doubles that are the
q-deformations  for Lie algebras or Lie superalgebras.
By studying its representation theory,many-parameter  representations
of the exotic quantum double are obtained with an explicit example
associated with Lie algebra $A_2$ .The multi-parameter R-matrices for
the quantum Yang-Baxter equation can result from the universal R-matrix
of this exotic quantum double and these representattions.

\newpage
\large
{\bf 1.Introduction}
\vspace{0.4cm}

In recent years,the quantum Yang-Baxter equation(QYBE)[1,2] has become a focus
of the attention from both theoretical physicists and mathematicians.This is
becuase the QYBE is a key to the complete integrability of many physical
systems  appearing in the quantum inverse scattering methods[3,4],the exactly-
solvable models in statistical mechanics[5] and low-dimensional quantum
field theory[6].In solving the QYBE in a general way and classifying
 its solutions
(R-matrices)algebraically,a remarkble mathematical structure-the
quasi-triangular Hopf algebra(loosely called quntum group) are
found in connection with
the QYBE[7-10].Among these developments,the Drinfeld's quantum double[7]
theory provides one with a general construction to systematically obtain
solutions of the QYBE in terms of the quantum doubles(QDs)
,which  usually are the `q-deformations'of certain algebras,
 and their representations.The recent studies show that,not only
the standard R-matrices[11-13],but also the non-standard ones[14-16],such as
the R-matrices with non-additive spectral parameters[17-23],the colored
R-matrices[24-28],can be obtained in the framework of Drinfeld's
QD theory,but for the latter the cyclic representations and other non-generic
ones at roots
of unity[30-39] and some paramterization of the quantum (universal eveloping)
algebras [40-41] need to be considered.The purpose of the present paper is
to search for the exotic quantum doubles ,other than those `q-deformations',
so that the new universal R-matrix can be obtained for the QYBE based
on Drinfeld's quantum double theory.

To proceed our discussion conveniently,we need to outline some basic ideas
in Drinfeld's QD theory so that the notations used in this paper can be
clarified.Suppose we are given two Hopf
algebras A ,B and a non-degenerate bilinear form $< , >$:
$A\times B\rightarrow C$(the complex field) satisfying the following
conditions:
$$<a, b_1 b_2>=<\Delta_A (a), b_1\otimes b_2>, a\in A,~b_1,~b_2\in B,$$
$$<a_1 a_2,b>=<a_2\otimes a_1,\Delta_B(b)>,a_1,a_2\in A,b\in A $$
$$<1_A,b>=\epsilon_B(b),b\in B,\eqno{(1.1)}$$
$$<a,1_B>=\epsilon_A(a),a\in A$$
$$<S_A(a),S_B(b)>=<a,b>,a\in A,b\in B$$
where for C=A,B,$\Delta_C,\epsilon_C$ and $S_C $ are the coproduct,counit and
antipode of C respectively;$1_C$ is the unit of C.Drinfeld's QD theory (for a
comprehensive reviews see the refs.[42,44]) states
the central
results in the QD theory as follows:
\vspace{0.4cm}

{\bf Theorem 1}.{\it There exists a Hopf algebra D satisfying the following
conditions\\
1.D contains A and B as Hopf subalgebras;\\
2.The mapping $A\times B\rightarrow D:a\otimes b\rightarrow ab$ is an
isomorphism of vector space;\\
3.For any $a\in A,b\in B$,we have multiplication
$$ba=\sum_{i,j} <a_i(1),S(b_j(1))><a_i(3),b_j(3)>a_i(2)b_j(2) \eqno{(1.2)}$$
where $c_i(k)(k=1,2,3;c=a,b)$ are defined by
$$\Delta^2(c)=(id\otimes\Delta)\Delta(c)=(\Delta\otimes id)\Delta(c)=
\sum_i c_i(1)\otimes c_i(2)\otimes c_i(3)$$}
\vspace{0.4cm}
{\bf Theorem 2} {\it There exists an unique element

$$\hat{R}=\sum_m a_m\otimes b_m \in A\times B\subset D\times D $$\\
obeying the ``abstract''QYBE\\
$$\hat{R}_{12}\hat{R}_{13}\hat{R}_{23}=\hat{R}_{23}\hat{R}_{13}\hat{R}_{12},
\eqno{(1.3)}$$
where $a_m$ and $b_m$ are the basis vecors of A and B respectively,and they
are dual each other,ie.,$<a_m,b_n>=\delta_{m,n}$;\\
$$\hat{R}_{12}=\sum_m a_m\otimes b_m\otimes 1,\\
\hat{R}_{13}=\sum_m a_m\otimes 1 \otimes b_m,\\
\hat{R}_{23}=\sum_m 1\otimes a_m\otimes b_m$$
where 1 is the unit of D.}

\vspace{0.4cm}

Up to now,the QD's built explicitly are only the quantum (universal eveloping)
algebras and superalgebras and their parameterizations.They are the
q-deformations of the universal
algebras and possess a `standard' quantum double structure that both
 the subalgebras A and B are non-commutative and non-cocommutative.This
symmetric structure reflects the duality of A and B.Notice that these
standard quantum doubles approach the usual universal enveloping algebras
(UEA)in the classical limit $q\rightarrow 1$.In this paper ,we will
consruct so-called exotic quantum doubles (EQD) that are not those
q-defomations and possess asymmetric dual structure that one of the
subalgebras A and B is commutative but non-cocommutative and another
cocommutative but non-commutative.As new quasi-triangular Hopf algebras,
these EQDs naturally enjoy the QYBE,but they have not the usual classical
limit.

This paper is arranged as follows.In section 2,we take the sub-Borel
subalgebra of the UEA of the classical Lie algebra as the Hopf subalgebra
A with cocommutative coproduct in the QD construction and then built
its quantum dual as a non-cocommutative but commutative Hopf subalgebra
B.In section 3,we combine A and B to form  the exotic quantum double
and thereby obtain the new universal R-matrix for the QYBE.In section 4,
we discuss an explicit example of EQDs ,which is connected with the
Lie algebra $A_2$ in details.In section 5,we study  the representation
theory of the EQD with the above example and construct a class of
many-parameter representations to built the many-parameter R-matrices
for the QYBE.Finally,in section 6,we give some remarks on the problems and
the possible developments in the EQD.

\vspace{0.5cm}
{\bf 2.Quantum Dual for Non-simple Lie Algebra}\\
 \vspace{0.4cm}

Let $\phi^+ :\{\alpha , \beta, \gamma...\}$ be the system of all positive
roots with respect to a simple root system  of a calassical Lie algebra L.
A Cartan elements $h\in$ H(the Cartan subalgebra and all the positive root
vectors $\{  e_{ \alpha},\mid \alpha \in \phi^+ \}$ generate an associative
algebra
A with the relations on the Cartan-Weyl basis\\
$$[h,e_{ \alpha}]=\alpha (h)e_{ \alpha},\eqno{(2.1)}$$
$$[e_{ \alpha},e_{ \beta}]=N_{\alpha, \beta}e_{ \alpha +\beta }$$
where $\alpha \in H^*$ and the coefficients  $N_{\alpha, \beta}$ enjoy the
structure of the Lie algebra L.In fact,the algebra A is a subalgebra of the
Borel subalgebra of the universal enveloping algebra (UEA) of the Lie algebra
L.
Defining the algebraic homomorphisms $\Delta:A
\rightarrow A\otimes A,\epsilon:A\rightarrow {\it C}$ and the algebraic
antihomomorphism $S:A \rightarrow A$ by
$$\Delta(x)=x\otimes 1+1\otimes x,S(x)=-x,\epsilon(x)=0\eqno{(2.2)}$$
where $x\in \{h,e_{ \alpha}\mid  \alpha \in \phi^+ \}$,one gives the algebra
a ``trivial''(cocommutative) Hopf algebraic structure.It is a well-known
fact in the theory of Hopf algebra since we can regarded the algrbra A
as an UEA of the non-simple Lie algebra
with basis $\{h,  e_{ \alpha},\alpha \in \phi^+ \}$.However ,the Hopf
algebraic dual(quantum dual) B of A is non-trivial(non-cocommutative)
due to the duality of B to A.Now,we derive the structure of A in tems of
this duality.

Becuase A is cocommutative,its dual is an Abelian algebra with commuting
generators.So the associative algebraic structure is quite simple.
To consider the Hopf algebraic structure ,
 we set an order for the basis of A:If $\alpha -\beta$ is a
no-zero positive root,then  we say $\alpha  \succ \beta$ ;the basis
for A is writen down to enjoy this order as
$$\{ a(m,m_{\alpha})=h^m \prod _{\alpha \in \phi^+} e^{m_ \alpha}_ \alpha
=h^m...e^{m_ \beta}_
\beta... e^{m_ \gamma}_ \gamma ...e^{m_ \delta}_\delta ...\mid$$
$$...\delta \succ ...\gamma...\succ ...\beta ,m_\alpha \in {\it Z^+}
=\{0,1,2,..\}\}$$
Suppose that the dual Hopf algebra B to A is generated by the
dual generators $t,f_\alpha( \alpha \in \phi^+)~to~ h,e_{ \alpha}$
respectively.They are defined by the following pairs
in terms of a bilinear form $<~,~>$
2~
$$<h,t>=1,<x,t>=0,<e_\alpha,f_\alpha>=1,<y,f_\alpha>=0;\eqno{(2.3)}$$
where x and y are the basis elents of A other than h ,$ e_{ \alpha}$
respectively.
\vspace{0.5cm}

{\bf Proposition 1}.{\it
For $m_\alpha,n_\alpha,m \in {\it Z^+}(\alpha \in \phi^+)$,
$$<h^m \prod _{\alpha \in \phi^+} e^{m_ \alpha}_ \alpha ,t^l
 \prod _{\alpha \in \phi^+} f^{n_ \alpha}_ \alpha>= \delta_{m,l}m!
 \prod _{\alpha \in \phi^+}m_\alpha !\delta_{m_\alpha,n_\alpha}, \eqno{(2.4)}$$
namely,the vectors
$$ b(m,m_\alpha)= \frac{t^m  }{m! }  \prod _{\alpha \in \phi^+}
\frac {f^{m_ \alpha}_ \alpha}{m_\alpha !},\eqno{(2.5)}$$
form a dual basis for B to $ a(m,m_{\alpha})$ respectively:
$$< a(m,m_{\alpha}), b(n,n_\alpha)>=\delta_{m,n}
 \prod _{\alpha \in \phi^+}\delta_{m_\alpha,n_\alpha}$$}
\vspace{0.5cm}

{\bf Proof}.Thanks the duality between A and B,we have
$$<h^l,t^m>=<\Delta(h^l),t^{m-1}\otimes t>= \sum_{k=0}^{l} \frac{l!}
{k!(l-k)!}<h^{l-k}\otimes h^k,t^{m-1}\otimes t>$$
$$=l<h^{l-1},t^{m-1}>=....=l!\delta_{l,m}$$
Similarly,for $G=e_{ \alpha},F=f_{ \alpha}$ respectively,
$$<G^m,F^n>=m!\delta{m,n}$$
Then,
$$<h^mG^n,F^s>= \sum_{k=0}^{m} \sum_{r=0}^{n}\frac{m!n!}{(m-k)!k!(n-r)!r!}
<h^{m-k}G^{n-r}\otimes h^kG^r,F\otimes F^{s-1}>$$
$$= \sum_{k=0}^{m} \sum_{r=0}^{n}\frac{m!n!}{(m-k)!k!(n-r)!r!}\delta_{n-r,1}
\delta_{m-k,0}<h^kG^r,F^{s-1}>$$
$$=n<h^mG^{n-1}.F^{s-1}>=n!\delta_{n,s}\delta_{m,0};$$
$$<h^mG^n,t^rF^s>= \sum_{k=0}^{m} \sum_{r=0}^{n}\frac{m!n!}{(m-k)!k!(n-r)!r!}
<h^{m-k}G^{n-r}\otimes h^kG^r,t^l\otimes F^s>$$
$$=<h^m,t^l><G^n,F^n>.$$
It follows from the above calculations that
$$<a(m,m^\alpha),b(n,n^\alpha)>=<h^m,t^n> \prod _{\alpha \in \phi^+}<
 e^{m_ \alpha}_ \alpha,f^{n_ \alpha}_ \alpha>=m!\delta_{m,n}
\prod _{\alpha \in \phi^+}m_\alpha !\delta_{m_\alpha ,n_ \alpha}$$
\vspace{0.5cm}

In this position,we can deduce the Hopf algrbraic structure,ie,$
(\Delta=\Delta_B,\epsilon=\epsilon_B ,S=S_B)$ of the algebra B.Let us
first consider $\Delta(f_{\gamma})$.
Notice that the reduced linear form $<~,\Delta (f_{\gamma})>$ is non-zero only
on $e_\gamma\otimes 1,h^n\otimes e_\gamma,h^ne_{ \alpha}\otimes  e_{ \beta}
( \beta\succ \alpha)$:
$$<e_\gamma\otimes 1,\Delta( f_{\gamma})>=<1.e_\gamma, f_{\gamma}>=1;$$
$$<h^n\otimes e_\gamma,\Delta( f_{\gamma})=< e_\gamma h^n, f_{\gamma}>
=<(h-\gamma(h))^ne_\gamma,f_{\gamma}>=(-\gamma(h))^n;$$
$$<h^ne_{ \alpha}\otimes  e_{ \beta}, f_{\gamma}>=<(h-\beta(h))^n
 e_{ \beta}e_{ \alpha},f_{\gamma}>$$
$$=(-\beta(h))^n<e_{ \alpha} e_{ \beta}-N_ { \alpha, \beta}e_{\alpha,\beta}
,f_{\gamma}>
=-(-\beta(h))^nN_ { \alpha, \beta}
\delta(\alpha+\beta,\delta).$$
Consequently,
$$\Delta( f_\gamma)= f_\gamma\otimes 1+ \sum_{n=0}^{\infty}
\frac{(-\gamma (h))^n t^n} {n!} \otimes  f_\gamma$$
$$- \sum _{\alpha,\beta \in \phi^+ }
N_ { \alpha, \beta}\delta(\alpha+\beta,\gamma)\theta(\beta-\alpha)
\sum_{n=0}^{\infty} \frac{(-\beta(h))^n t^n}{n!}f_{ \alpha}\otimes
f_\beta$$
where
$$\delta(\alpha,\beta)=\left\{ \begin{array}{ll}
1, &\mbox{if $\alpha =\beta;$}\\
0, &\mbox{if $\alpha \neq \beta$}
\end{array}
\right.$$

$$\theta(\alpha)=\left\{ \begin{array}{ll}
1, &\mbox{if $ \alpha (\neq 0) \in \phi^+ ;$}\\
0, &\mbox{if $\alpha \not\in   \phi^+ $ }
\end{array}
\right.$$

For $S(f_\gamma)$ and $\beta \succ \alpha$,the only non-zero pairs are
$$<h^ne_\alpha  e_\beta,S(f_\gamma)>=<(-1)^nS(e_\beta  e_\alpha h^n )
,S(f_\gamma)>=(-1)^n<e_\beta  e_\alpha h^n,f_\gamma>$$
$$=(-1)^n<(h-\beta(h)-\alpha(h))^n(e_\alpha e_\beta-N_ {\alpha,\beta}
e_ {\alpha+\beta}),f_\gamma>
=-(\beta(h)+\alpha(h))^nN_ {\alpha,\beta}\delta_{\alpha+\beta,\gamma};$$
$$<h^n e_\gamma,S(f_\gamma)>=-(-1)^n<S (e_\gamma h^n),S(f_\gamma)>
=-(-1)^n<e_\gamma h^n,f_\gamma>$$
$$=-(-1)^n<(h-\gamma(h))^ne_\gamma ,f_\gamma>=-(\gamma(h))^n$$
Consquently,
$$S(f_\gamma)= -\sum_{n=0}^{\infty}\frac{\gamma(h)^nt^n}{n!}
f_\gamma$$
$$+\sum_{\alpha,\beta \in \phi^+}  N_ {\alpha,\beta} \theta(\beta-\alpha)
\delta(\alpha+\beta,\gamma)\sum_{n=0}^{\infty}
\frac{(\beta(h)+\alpha(h))^nt^n}{n!}f_ \alpha  f_\beta $$
In the same way we derive $\Delta(h),S(h),\epsilon(f_ \alpha)$ and so on.
The results are listed as follows.
\vspace{0.5cm}

{\bf Proposition 2}.{\it ~The duality between A and B results in the
the commutative associative algebraic structure and the non-cocomutative
Hopf algebraic structure defined by
$$\Delta(t)=t\otimes 1+1\otimes t$$
$$\Delta(f_\gamma)=f_\gamma\otimes 1+e^{-\gamma(h)t}\otimes f_{\gamma}-
\sum _{\alpha ,\beta \in \phi^+ }C(\alpha,\beta ,\gamma )
e^{-\beta(h)t}f_\alpha\otimes f_\beta,$$
$$S(f_\gamma)=-e^{\gamma(h)}(f_\gamma+\sum_{\alpha ,\beta \in \phi^+ }
C(\alpha,\beta ,\gamma )f_\alpha f_\beta),  \eqno{(2.6)}$$
$$S(h)=-e^{-\gamma(h)t}h,S(1)=1;\epsilon( f_\gamma)=\epsilon(h)=0,
\epsilon(1)=1.$$
where
$$C(\alpha,\beta,\gamma)= N_ {\alpha,\beta}
\theta(\beta-\alpha)\delta(\alpha+\beta,\gamma).$$}

\vspace{0.5cm}
{\bf 3.The Quantum Double and Its Universal R-Matrix}
 \vspace{0.4cm}

In this section we show how the algebra A and its quantum daul B can
be combined to form a quasi-triangular Hopf algebra with the exotic structure.
To define the multiplications between A and B,we need to use the following
formula
$$\Delta^2(x)=x\otimes 1\otimes 1+1\otimes x \otimes 1+1\otimes 1\otimes x
,x=h,t,e_\alpha,\alpha \in \phi^+$$
$$\Delta^2( f_\gamma)= f_\gamma\otimes 1\otimes 1+e^{-\gamma(h)t}\otimes
 f_\gamma\otimes 1+e^{-\gamma(h)t}\otimes e^{-\gamma(h)t}\otimes f_\gamma$$

$$-\sum_{\alpha ,\beta \in \phi^+ }C(\alpha,\beta ,\gamma )e^{-\gamma(h)t}
\otimes e^{-\beta(h)t}f_ \alpha\otimes f_\beta-\sum_{\alpha ,\beta \in \phi^+}
C(\alpha,\beta ,\gamma )e^{-\beta(h)t}f_ \alpha\otimes f_\beta\otimes 1$$

$$-\sum_{\alpha ,\beta \in \phi^+ }C(\alpha,\beta ,\gamma )e^{-\beta(h)t}
f_ \alpha\otimes e^{-\beta(h)t}\otimes f_\beta-$$
$$\sum_{\alpha ,\beta,\sigma,
\delta \in \phi^+ }C(\alpha,\beta ,\gamma)C(\sigma,\delta,\beta)
e^{-\beta(h)t}f_ \alpha\otimes e^{-\delta(h)t}f_\sigma\otimes f_\delta,
\eqno{(3.1)}$$
Using the above equations and the defination (1.2),we calculate the commutators
$[e_\alpha, f_\gamma],[h, f_\gamma],[t,e_\alpha]$ and so on:

$$ f_\gamma e_\gamma=< e_\gamma,S( f_\gamma)><1,1>1.1+$$
$$<1,S(e^{-\gamma(h)t})><1,1> e_\gamma f_\gamma+<1,S(e^{-\gamma(h)t})>
<e_\gamma f_\gamma>1.e^{-\gamma(h)t}$$
$$=-1+e_\gamma f_\gamma+e^{-\gamma(h)t};$$
$$f_\gamma e_\eta=-\sum_{\alpha ,\beta \in \phi^+ }C(\alpha,\beta ,\gamma )
<e_\eta,S(e^{-\beta(h)t}f_ \alpha)><1,1>1.f_\beta+<1,S(e^{-\gamma(h)t} )><1,1>
e_\eta f_\gamma$$
$$-\sum_{\alpha ,\beta \in \phi^+ }C(\alpha,\beta ,\gamma )
<1,S(e^{-\gamma(h)t})><e_\eta,f_\beta>e^{-\beta(h)t}   f_\alpha$$
$$=\sum_{\beta \in \phi^+ }C(\eta,\beta ,\gamma )f_\beta
-\sum_{\alpha \in \phi^+ }C(\alpha,\eta ,\gamma )
e^{-\eta(h)t}   f_\alpha+e_\eta f_\gamma;   \eqno{(3.2)}   $$
$$ f_\gamma h=<h,,S(e^{-\gamma(h)t}><1,1> f_\gamma+$$
$$<1,,S(e^{-\gamma(h)t}><1,1>h f_\gamma
+\gamma(h) f_\gamma+h f_\gamma;$$
$$tx=<1,1><1,1>xt$$
The above results are rewritten as follows
\vspace{0.4cm}

{\bf Proposition 4}.{\it The multiplication between A and B is defined by the
 following commutators
$$[e_\alpha,f_ \alpha]=1-e^{-\alpha(h)t},$$
$$[h,f_ \alpha]=-\alpha(h)f_\alpha,\eqno{(3.3)}$$
$$[e_\alpha,f_\beta]=\sum_{\gamma\in \phi^+ }C(\gamma,\alpha,\beta )
e^{-\alpha(h)t}f_\gamma-\sum_{\gamma\in \phi^+ }C(\alpha,\gamma ,\beta )
f_\gamma,$$
$$\alpha \neq \beta,$$
$$[t,x]=0,x=h,e_\alpha$$}
\vspace{0.4cm}

The above commutators combine the algebra A with  its quantum dual B
to form a non-cocommutative and non-commutative Hopf algebra D(A)=D
 as the quantum
double of A(or B).As an associative algebra ,it is generated by $h,t,
e_\alpha ,f_ \alpha,(\alpha \in \phi^+) $and the unit 1 obeying eqs.(2.1)
,and endowed with the
Hopf algebraic structure by eqs.(2.2) and (2.6).Now,let us show that
this Hopf algebra D is also quasi-triangular.In fact,the construction of
Drinfeld's QD theory automatically perseveres the existence of the
quasi-triangular stucture.Intertweening A and B,the universal
R-matricx is a canonical element
$$\hat{R}=\sum_{m,m_\alpha =0,\alpha \in \phi^+}^{\infty }
a(m,m_\alpha)\otimes b(m,m_\alpha ) =e^{h\otimes t}
 \prod _{\alpha \in \phi^+ }exp(e_\alpha \otimes f_\alpha). \eqno{(3.4)}$$
This element $\hat{R}  (\in D\otimes D)$ endows the Hopf algebra D with
a quasi-triangular stucture enjoyed by the following relations
$$\hat{R}\Delta(x)=\sigma\Delta(x)\hat{R},$$
$$(\Delta\otimes id)\hat{R}=\hat{R}_{13}\hat{R}_{23},$$
$$(id\otimes\Delta)\hat{R}=\hat{R}_{13}\hat{R}_{12}, \eqno{(3.5)}  $$
$$(\epsilon\otimes id)\hat{R}=1=(id\otimes\epsilon)\hat{R},$$
$$(S\otimes id)\hat{R}=\hat{R}^{-1}=(id\otimes S)\hat{R},$$
where $\sigma$ is such a permutation that $\sigma(x\otimes y)=y\otimes x
,x,y\in D.$The eqs.(3.5) imply that the above constructed universal R-matrix
satisfies the abstract QYBE.It is not too difficult to verify the
above relations (3.5) by a straightforward calculation.

In the above discussion,we have constructed a new quantum ``group''( quasi-
triangular Hopf algebra) D associated with an arbitrary classical Lie algebra
in terms of Drinfeld's QD theory.In comparision with the ``standard''
 quantum `groups' that are the q-defomations of UEA's of classical Lie
algbras and superalgebras,our  quantum `group' D possesses some new
features:1.D has not the usual classical limit since it is not a
q-deformation of the QEA.2.It has an exotic subalgebraic structure that
the subalgebra A is cocomutative but not commutative and the subalgebra B
commutative but not cocommutative.This asymmetric structure is quite
different from the symmatric structure that both A and B are non-commutative
and non-cocomutative.We will call D exotic quantum double.\\

\vspace{0.5cm}
{\bf 4.Example of The Exotic Quantum Double for $A_2$}
\vspace{0.4cm}

In this section an explicit example of the exotic quantum double will be
given in connection with the classical Lie algebra $A_2$.In this example.the
subalgebra A is taken to be an associative algebra generated by $h,a,b$
 and the
relations
$$[h,a]=\mu a,[h,b]=b, \eqno{(4.1a)}$$
$$[a,[a,b]]=0=[b,[b,a]], \eqno{(4.1b)}$$
The generators a and b can be regarded as the root vectors with respect to
the simple roots $\alpha_1$ and $\alpha_2$ respectively for $A_2$.
The third positive root vector corresponding to  $\alpha_1+\alpha_2$
is just the commutator of a and b,ie.,
$$c=[a,b]   ,\eqno{(4.2)}$$
which satisfies
$$[c,a]=0=[c,b],$$
$$[h,c]=(\mu+1)c.\eqno{(4.3)}$$
The first equation in eq.(4.3) results from the Serre relation (4.1b).If
we take $h_1$ and $h_2 $ as the Cartan elements in the Chevalley basis for
 $A_2$ and
$$[h_1,a]=2a,[h_1,b]=-b,$$
$$[h_2,a]=-a,[h_2,b]=2b,\eqno{(4.4)}$$
then,
$$h=\frac{2\mu+1}{3}h_1+\frac{\mu+2}{3}h_2. \eqno{(4.5)}$$
The cocomutative Hopf algebraic structure of A is endowed with by
$$\Delta(x)=x\otimes 1+1\otimes x$$
$$S(x)=-x,S(1)=1;\epsilon(x)=0,\epsilon(1)=1$$

Let B  be the quantum daul to A and $t,d,f,g$ be its dual generators to
$h,a,b,c$ respectively.According to the last sections,a straightforward
calculation gives the Hopf algebraic structure of B:
$$\Delta(d)= d\otimes 1+e^{-\mu t}\otimes d,$$
$$\Delta(f)=f\otimes 1+e^{-t}\otimes f,$$
$$\Delta(g)=g\otimes 1+e^{-(\mu+1) t}\otimes g-e^{-t}d\otimes f,$$
$$S(d)=-e^{\mu t}d,$$
$$S(g)=-e^{(\mu+1) t}(g+d\otimes f),$$
$$S(f)=e^{t}f, \eqno{(4.7)}$$
and the multiplication relation between A and B
$$[h,d]=-\mu d,$$
$$[h,f]=-f,$$
$$[h,g]=-(\mu +1)g,$$
$$[a,d]=1-e^{-\mu t},$$
$$[b,f]=1-e^{-t},$$
$$[c,g]=1-e^{-(\mu+1)t},\eqno{(4.8)}$$
$$[a,g]=-f,$$
$$[b,g]=e^{-t}d,$$
$$[a,f]=0=[b,d],$$
$$[c,d]=0=[c,f],$$
$$[t,s]=0,x=a,b,c,h.$$
The quantum double D(2) is generated by $a,b,c,h,t,d,e,f$ with the relations
(4.1a),(4.2) (4.3) and (4.8) as an associative algebra.Its quasi-triangular
Hopf algebraic stucture is endowed with by eqs.(4.7) and the universal R-matrix
$$\hat{R}=e^{h\otimes t}e^{a\otimes d}e^{h\otimes f}e^{c\otimes g}
 ,\eqno{(4.9)}$$

\vspace{0.5cm}
{\bf 5 The Representation Theory and Many-Parameter R-matrices}
\vspace{0.4cm}

One purpose of building  quantum double is to obtain the solutions
of the QYBE in terms of its universal R-matrix and matrix representations.
In order to find the solutions  of QYBE associated with the
exotic quantum double D,we should study its representation theory.In fact,
for a given repreentation $T^{[x]}$ of D:
$$T^{[x]} :D\rightarrow End(V)$$
on the linear space V where x is a continuous parameter,we can construct
a R-matrix
$$ R(x,y)=T^{[x]}\otimes  T^{[y]} (\hat{R})$$
satisfying the QYBE
$$R_{1,2}(x,y)R_{1,3}(x,z)R_{2,3}(y,z)=R_{2,3}(y,z)R_{1,3}(x,z)R_{1,2}(x,y),
\eqno{(5.1)}$$
Here,x,y and  z appear as the color parameters [24] similar to the
non-additive spectrum parameters in QYBE.
This additivity for R-matrices was first found in ref.[15,16] for the
chiral Potts model in statistical mechanics.Thus,it is necessary to study
the representation theory and construct the many-parameter representations
for the exotic quantum double D.However,to write down an explicit
representation of a general D is rather overlaborate.So we only discuss
the typical example D(2) in this section,but the main ideas and method can be
directly applied to the general case.

To simplify our discussion,we have to distinguish between the trivial and non-
trivial D- modules.
\vspace{0.4cm}

{\bf Defination 1}.{\it The action of an  operator on  the reprsentatrion space
V is called  to be  trivial if its kernal is the whole space that it acts on.}
\vspace{0.4cm}

{\bf Defination 2}.{\it A D(2)-module V is called to be trivial if at least
one of the generators of D acts trivially on V;otherwise,it is called
non-trivial module.}\\

Before studying the representation theory of D(2),we would like to give a
remark on the above definations.To study the  trival D(2)-mdule is much easier
than that of a non-trivial one.In fact,the structure of a non-trivial D(2)
module collapses into that of the module of a simpler algebra $D'$.For
example,if the action of t in D(2) is trivial,one only need to study the module
of
the asssociative algebra generated by $h,a,b,c,d,f,g$ with non-zero commutatiom
relations
$$[a,b]=c,[a,g]=-f,[b,g]=d $$
$$[h,a]=\mu a,[h,b]=b,$$
For this reason,we will mainly study the non-trivial D(2)- model.

Having the above description,we are now  in the position to prove a
proposition as a central result for the representation theory of D(2).

\vspace{0.7cm}
{\bf Proposition 5}.{\it There doese not exist a finite dimensional
irreducible  D(2)-module.}

\vspace{0.4cm}

{\bf Proof.}Suppose there exists a  finite dimensional irreducible  D(2)-model
V and $T:D\rightarrow End(V) $ is the corresponding  finite dimensional
irreducible  representation.For simplicity we by x denote T(x) as follows for
$x
\in D(2)$.Since t belongs to the center of D,t must be a non-zero scalar in
 non-trivial finite dimensional irreducible representation according to
the Schur lemma.Otherwise,if t is zero,V is trivial.Because
$ {\it C}$ is an algebraic closure,there must be an eigenvector v such that
$$ hv=\xi v,\xi \in {\it C}$$
Noticing the vectors $ v,av,a^2v,...,a^n v...$ corespond to the distinct
eigenvalues $\xi,\xi+\mu,\xi+2\mu,...,\xi+n\mu....$ for $\mu\neq 0$ ,we
come to the conclusion that there exists $r\in  {\it Z^+}$ such that
$ v,av,a^2v,...,a^{r-2 }v, a^{r-1}v=u$  are linearly independent and $a^nv
=0$ .Similarly,there are $s,q\in  {\it Z^+} $ such that rhe non-zero
$u,bu,b^2u,...,
b^{s-1}u=w$ are  are linearly independent and $bw=b^su=0$;the non-zero vectors
$w,cw,c^2w,....,
c^{q-1}w =z $ are linearly independent and $cz=c^qw=0$.Then,we can prove that
$az=bz=cz=0$
and thus the vector z generates a D(2)-submodule
$${\it S}=Span\{F(m,n,l)=d^mf^ng^l z\mid m,n,l \in  {\it Z^+}\}$$
under the action of D(2).Thanks to the irreducibility of V and its finite
dimension,we must have $ {\it S}  =V$ and conclude that there must exist $
m',n'
,l'$ so that
$$d F(m'-1,n,l)        =0,$$
$$f F(m,n'-1,l)        =0, \eqno{(5.2)}$$
$$g F(m,n,l'-1)        =0,$$
that is to say,the dimension of V is  $m'n'l'$ of ${\it S} $.
However,it follows from eq.(5.1) that
$$ 0=ad F(m'-1,0,0)=ad^{m'}z$$
$$=[d^{m'}a+m'(1-e^{-\mu t})]z=m'(1-e^{-\mu t})z,$$
that is,$m'=0$.Similarly,$n'=l'=0$.This means the D(2)-module is trivial.
\vspace{0.4cm}

According to the above proposition,for the study of non-trivial representation
,we only need to ffocus on two cases,the indecomposable(reducible,but not
completely reducible) representations and the infinite dimensioanal irreducible
representations.Now,we only discuss the later .To construct an infinite
dimensional irreducible representation explicitly,we define a Verma-like
space
$$ V(\eta,\pi)=Span\{\mid M>=\mid m,n,l>=a^mb^nc^l\mid 0(\eta,\pi)>\mid
m,n,l \in {\it Z^+}\}$$
based on the vacuum-like state $\mid 0(\eta,\pi)>$:
$$d\mid 0(\eta,\pi)>=f\mid 0(\eta,\pi)>=g\mid 0(\eta,\pi)>=0$$
$$h\mid 0(\eta,\pi)>=\eta \mid 0(\eta,\pi)>,t\mid 0(\eta,\pi)>=
\pi\mid 0(\eta,\pi)>, \eqno{(5.3)}$$
where $\eta,\pi \in  {\it C}$.The existence of the vacuum-like state
$\mid 0(\eta,\pi)>$ is easily proved by considering that t and h commuts
each other and $(a,b,c),(d,f,g)$ and $(h,t)$ act as the `lifting' operators
,`lowing ' operator and the Cartan operators respectively  for a classical
Lie algebra.

\vspace{0.4cm}
{\bf Proposition 6}.{\it On the Verma-like space,the infinite dimensional
repreentation $T^{[\eta,\pi]}$
$$h\mid M >=[\eta+m\mu+(1+\mu)l]\mid M >,$$
$$a\mid M >=\mid M+e^1 >$$
$$b\mid M >=\mid M+e^2 >-m\mid M-e^1+e^3 >$$
$$c\mid M >=\mid M+e^3 >$$
$$d\mid M >=m(e^{-\mu\pi}-1)\mid M-e^1 >, \eqno{(5.4)}$$
$$f\mid M >=n(e^{-\pi}-1)\mid M-e^2 >$$
$$g\mid M >=l(e^{-(\mu+1)\pi}-1)\mid M-e^3 >+mn(e^{-\pi}-1)\mid M-e^1
-e^2>$$
is irreducible .Here $e^1=(1,0,0),e^2=(0,1,0),e^3=(0,0,1)$ are  the unit
vectors in the lattice sapce $ {\it Z^3} :\{M=(m,n,l)\mid m,n,l\in
 {\it Z^+}\}  $}
\vspace{0.4cm}

{\bf Proof}.Using the commutation relations of D(2),we can first prove
by induction for $n\in  {\it Z^+}$
$$ da^n=a^nd+n(e^{-\mu t}-1)a^{n-1},$$
$$ fb^n=b^nf+n(e^{- t}-1)b^{n-1},$$
$$ gc^n=c^ng+n(e^{-(\mu+1) t}-1)c^{n-1},$$
$$ga^n=a^ng+nfa^{n-1},$$
$$gb^n=b^ng-ne^{- t}db^{n-1} , \eqno{(5.5)}  $$
$$ba^n=a^nb-nca^{n-1},$$
$$ab^n=b^na+ncb^{n-1},$$
$$ha^n=a^nh+n\mu a^n,$$
$$hb^n=b^nh+nb^n,$$
$$hc^n=c^nh+n(1+\mu)b^n.$$
The eqs.(5.4) follows from eqs.(5.5 )and (5.3) immediately.It is not
difficult to verify that the eqs.(5.4) indeed define a representation of D(2).
By considering that the indices m,n and l not only decrease but also
increase by unit 1,it can be proved that this representation is irreducible
if it is non-trivial.
\vspace{0.4cm}

Let us make an observation that there exist many parameter $\mu,\pi~ and
{}~\eta.$ Among them $\mu$ and $\eta$ are allowed by the quantum double
structure and the representation theory respectively while $\pi$ is due to
the existence the central element t.Since $\eta$ and $\pi$ can be used to
distinguish
the different representations ,we can set $x=(\eta,\pi )$ and
obtain the colored R-matrices
with two dimensional color parameters x where the parameter  $\mu$ is intrinsic
and plays the similar role to that of  the q in the standard quantum double
-quantum algebras.In fact,for a rank-$l$ classical Lie algebra,we can
introduce $l-1$ independent intrinsic paramters to the corresponding
exotic quantum double since its Cartan subalgebra is   $l$ dimensional.

\vspace{0.5cm}
{\bf 6.Discussions}
\vspace{0.7cm}

To conclude this paper,we should give some remarks on our exotic quatum double
and its relations to the known results ,such as the Hopf algebraic structure
for the function algebra on the formal group [43,44 ],the extended
Hesenberg-Weyl
algebra (the boson algebra) as a quantum double [ 45-47] and so on.

\vspace{0.5cm}
{\bf i.)} From the construction of the exotic quantum double in this paper,
we can see that a commutative(Abelian) algebra ,eg.,the subalgebra B,can be
endowed with a non-cocommutative Hopf algebraic structure and its quatum
daul and quantum double can be deduced as  non-commutative algebras.
Such a process  can be
regarded as the inversion of the construction in this paper and maybe provide
us a scheme of `quantization' from commtative object to non-commutative
one.An example of this `quantization' was given [45 ] recently.A simplest
associative algebra is generated by two commuting generators X and H .Its
non-cocommutative Hopf algebraic structure is defined by
$$\Delta(H)=H\otimes 1+1\otimes H,\Delta(X)=X\otimes 1+e^{-H}\otimes X,$$
$$S(H)=-H,S(X)=-e^H X,\epsilon(X)=0=\epsilon(H)$$
Let Y and N be the daul generators to X and H respectively.Then the quantum
daul has cocommutative Hopf algebraic structure  the elements X,Y,H and N
generate a quantum double D(1) with the only non-zero commutation relations
$$[N,X]=X,[N,Y]=-Y,[X,Y]=1-e^{-H}$$
This quantum double D(1) is just the special example that A is the `Half' UEA
of $A_1$.There exists a homomorphisim
$$a\rightarrow X,a^+\rightarrow Y,E\rightarrow 1-e^{-H}, \hat N\rightarrow
-N$$
from the boson algebra generated by
the creation operator $a^+$ ,the annihilation operator $a$ ,the number
operator $\hat N$ and the central operator E to this quantum double where
the only non-zero commutation relations for the boson algebra are
$$[a, a^+]=E,[ \hat N,a]=-a,[\hat N,a^+]=a^+ $$
This example shows the so-called `quatization' from commtative object to
non-commutative
one in which the  quantum Yang-Baxter equation is enjoyed by the universal
R-matrix
$$\hat{R}=exp(X\otimes Y)exp(N\otimes H)$$

\vspace{0.4cm}
{\bf ii.)}It has to be pointed out that there are some difficulties in
the futher developments of the exotic quantum double theory.When one take
the subalgebra B to be the whole UEA of a classical algebra ,we hardly
write down the dual basis explicitly and so the construction scheme
of this paper can not work well.The similar problem also appears in
the discussion in terms of the formal group.How to generalize the method and
ideas of this paper to work on the case of the whole UEA other than a
Borel subalgebra is the first open question we should mention.The second
open question is how to find a finite dimensional representation for
the exotic quantum double except the example for $A_1$ mentioned above.
It is well-known that the  finite dimensional R-matrices usually make
sense in the quantum inverse scattering method and even in the exactly-
solvable models in stitistical mechanics.Thus,it is also expected that
some new  finite dimensional R-matrices can follow from the exotic quantum
double through its universal R-matrix where the finite dimensional
representations of the exotic quantum double must be used.However,although we
can do it
for the special case of $A_1$ by building the  finite dimensional idecomposable
representation of D(1) on certain quotient space on linear space D(1),
we can not obtain a finite dimensional representation for other higher-rank
exotic quantum double by the same method due to the existance of the
multi-center in cetain subalgebra.Therefore,there needs the futher works
on  finite dimensional representation.

\vspace{0.5cm}

{\bf iii.)}In the formal group theory of Lie algbera [43],the bialgebra
structure of the daul to the UEA of a classical Lie algebra can be given
abstractly in terms of the formal group.It is not difficult to further
define  the antipode for this dual bialgebra.So,in this abstract way,
the Hopf algebraic structure can be endowed with to this dual algebra.
However,writting out the explicit  Hopf algebraic structure,namely,
the the explicit multiplication relations ,  coproduct , antipode
and  counit for the dual generators,completely depends on the explict
evolution of the Baker-Comppell-Hausdorff formula for classical Lie
algebra.However,it is much difficult to do it even for the simple case
e.g.,SU(2).The study in this paper avoids this evalution so that not
only the dual Hopf algebraic structure is obtained,but also the corresponding
quantum double -the exotic quatum double is built
 for the Borel subalgebra of the UEA of
arbitrary classical Lie algebra by combining the two
subalgebras dual to each other.
\vspace{0.8cm}

\huge
Acknowledgements
\vspace{0.7cm}

\large
The author would like  to express his sincere thanks to
Prof.C.N.Yang for drawing his
attentions to the research field related to
the quantum Yang-Baxter equantion and for giving him very kindly helps.
He  thanks Prof.L.Tankhtajian for many useful and instructive discussions
who point out the possible relation of this work to the rlevant topics of
formal group.He also thanks Prof.B.McCoy for telling the works on chiral
Potts model and some discussions.
He is supported by Cha Chi Ming fellowship through the CEEC in State
University of New York at Stony Brook,and also is supported in part by the
NFS of China through Northeast Normal University .

\newpage

\huge
References\\

\large

\noindent
1.C.N.Yang,Phys.Rev.Lett.19(1967)1312\\
2.R.Baxter,Ann.Phys.70(1972)193\\
3.E.K.Sklyanin,L.A.Takhtajian and L.D.Faddeev,Theor.Mathem.Fisica 40(1979)194\\
4.P.P.Kulish and N.Y.Reshetikhin,J.Phys.A.16(1983),L591\\
5.R.J.Baxter,{\it Exactly-Solved Models in Statistical Machanics}
Academic.Press
,1982.\\
6.A.B.Zamolodchikov,Al.B.Zamolodchikov ,Ann.Phys.120(1979)253\\
7.V.G.Drinfeld,Proc.ICM.Berkeley,1986,(ed.By A.Gleason,AMS,1987),p.798\\
8.M.Jimbo,Lett.Math.Phys.10(1985)63\\
9.P.P.Kulish and N.Y.Reshetikhin,Zap.nauchn.Seminarod LOMI 101(1981)101\\
10. L.D.Faddeev and L.A.Takhtajian,Lect.Notice in Physics 246(1986)166,
Springer-Verlag\\
11.N.Y.Reshetikhin,LOMI preprints E-4 and E-11(1987)\\
12.M.Rosso,Commun.Math.Phys.117(1989)307\\
13.A.Kirillov and N.Y.Reshetikhin,Commun.Math.Phys.134(1990)421\\
14.H.C.Lee,M.Couture and N.C.Schmeing,Chalk River preprint CRNL-TP-1125,1988.\\
15.M.L.Ge,Y.S.Wu and K.Xue,Inter.J.Mod.Phys.A6(1991) 1645\\
16.Y.Cheng,M.L.Ge and K.Xue,Commun.Math.Phys.136(1991)196.\\
17.H.Au-Yang,B.McCoy,J.Perk,S.Tan and M.L.Yan,Phys.Lett.A123(1987)219\\
18.B.McCoy,J.Perk,S.Tan and C.H.Sah,Phys.Lett.A125(1987)9\\
19.R.J.Baxter,J.Perk,H.Au-Yang,Phys.Lett.A.128(1988)138\\
20.V.V.Bazhanov and Y.G.Stroganov,J.Stat.Phys.59(1990)799\\
21.E.Date,M.Jimbo,K.Mike and T.Miwa,Phys.Lett.A148(1990)45\\
22.E.Date,M.Jimbo,K.Mike and T.Miwa,Commun.Math.Phys.137(1991)133\\
23.V.V.Bazhanov ,R.M.Kashaev , V.V.Mangazeev and Y.G.Stroganov,Commun.Math.\\
Phys.138(1991)393\\
24.J.Murakami,Osaka preprints,1990\\
25.C.P.Sun,X.F.Liu and M.L.Ge,J.Math.Phys.32(1991)2409\\
26. M.L.Ge,X.F.Liu and C.P.Sun,Phys.Lett.A155(1991)137\\
27.M.L.Ge,X.F.Liu and C.P.Sun,Phys.Lett.A160(1991)433\\
28.M.L.Ge,C.P.Sun and K.Xue,,Inter.J.Mod.Phys.A7(1992)6609\\
30.G.Luatig,Contemp.Math.82(1989)59\\
32.P.Roche and D.Arnaudon,Lett.Math.Phys.17(1989)295\\
33.C.P.Sun,J.F.Lu and M.L.Ge ,J.Phys.A23(1990)L1199\\
34.L.Alvarez-Gaume,C.Gomez and G. Sirra,Nucl.Phys.B(1990)347\\
35.V.Pasquier and H.Saleur,,Nucl.Phys.B(1990)523\\
36.C.De Concini and V.G.Kac,{\it Representations of Quantun Group at
roots of 1},preprint 1990\\
37.C.P.Sun and M.L.Ge, ,J.Phys.A24(1991)L969;3265;3731.\\
38..C.P.Sun and M.L.Ge, ,J.Phys.A25(1992)19\\
39.C.P.Sun,H.C.Fu and M.L.Ge,Lett.Math.Phys.23(1990)19\\
40.X.F.Liu and C.P.Sun,Science in China A35(1992)73\\
41.X.F.Liu and M.L.Ge,Lett.Math.Phys. (1992)197\\
42.M.Jimbo,{\it The Topics from the Representations of $U(g)_q$     },in
Nankai Lect.on Math.Phys.ed by M.L.Ge,World Scientific,1992\\
43.J-P Serre,{\it Lie algebras and Lie Groups },V.A.Benjamin.INY,1965\\
44.L.A.Tankhtajian,{\it Quantum Groups},in Nankai Lect.Series In Mathematical
Phys.1989,ed by M.L.Ge and B.H.Zhao,World Scientific,1991\\
45.C.P.Sun,X.F.Liu,and M.L.Ge,J.Math.Phys.34(1992),in press\\
46.C.P.Sun,preprint ITP.SB-92-61,1992.\\
47.W.Li,C.P.Sun and M.L.Ge, ITP.SB-92-59,1992.\\

\end{document}